\documentclass[conference]{IEEEtran}
\IEEEoverridecommandlockouts
\usepackage{cite}
\usepackage{amsmath,amssymb,amsfonts}
\usepackage{algorithmic}
\usepackage{graphicx}
\usepackage{textcomp}
\usepackage{xcolor}
\usepackage{array}
\def\BibTeX{{\rm B\kern-.05em{\sc i\kern-.025em b}\kern-.08em
    T\kern-.1667em\lower.7ex\hbox{E}\kern-.125emX}}
\begin{document}

\title{Enhancing API Documentation through BERTopic Modeling and Summarization}

\author{\IEEEauthorblockN{}
\IEEEauthorblockA{\textit{}}}

\author{
\IEEEauthorblockN{AmirHossein Naghshzan}
\IEEEauthorblockA{\textit{École de technologie supérieure}\\
Montreal, Canada \\
amirhossein.naghshzan.1@ens.etsmtl.ca}
\and
\IEEEauthorblockN{Sylvie Ratté}
\IEEEauthorblockA{\textit{École de technologie supérieure}\\
Montreal, Canada \\
sylvie.ratte@etsmtl.ca}
}

\maketitle

\begin{abstract}

As the amount of textual data in various fields, including software development, continues to grow, there is a pressing demand for efficient and effective extraction and presentation of meaningful insights. This paper presents a unique approach to address this need, focusing on the complexities of interpreting Application Programming Interface (API) documentation. While official API documentation serves as a primary source of information for developers, it can often be extensive and lacks user-friendliness. In light of this, developers frequently resort to unofficial sources like Stack Overflow and GitHub. Our novel approach employs the strengths of BERTopic for topic modeling and Natural Language Processing (NLP) to automatically generate summaries of API documentation, thereby creating a more efficient method for developers to extract the information they need. The produced summaries and topics are evaluated based on their performance, coherence, and interoperability. 

The findings of this research contribute to the field of API documentation analysis by providing insights into recurring topics, identifying common issues, and generating potential solutions. By improving the accessibility and efficiency of API documentation comprehension, our work aims to enhance the software development process and empower developers with practical tools for navigating complex APIs.
\end{abstract}

\begin{IEEEkeywords}
Topic Modeling, Summarization, Natural Language Processing, LLM, BERTopic
\end{IEEEkeywords}

\section{Introduction}
The increasing availability of textual data across numerous fields, including software development, presents both challenges and opportunities for efficient information extraction. This expansion of data necessitates innovative methods for deciphering and presenting key insights, particularly from intricate resources such as Application Programming Interface (API) documentation. API documentation is a crucial asset for programmers, outlining how different software applications should interact. However, these documents can often be extensive, complex, and sometimes incomplete, creating hurdles in swiftly extracting pertinent information. Moreover, while official documentation serves as the primary source of information, previous studies indicate that it might not always be the optimal way to glean insights due to its voluminous and time-consuming nature \cite{Ponzanelli}.

These complexities often drive developers toward unofficial documents such as Stack Overflow and GitHub for more accessible information, highlighting the critical need for a more user-friendly approach to information extraction from API documentation. To address this need, our research introduces a novel method for the automatic generation of API summaries from informal documents like Stack Overflow. We leverage the power of BERTopic \cite{bertopic}, a topic modeling technique combining BERT (Bidirectional Encoder Representations from Transformers) \cite{bert} embeddings and c-TF-IDF (Class-based Term Frequency-Inverse Document Frequency), for extracting discussed topics. We also utilize Natural Language Processing (NLP) for text summarization, focusing on the Android programming language and its APIs and methods for this study. Our methodology begins with data collection and preprocessing, followed by the application of BERTopic modeling to extract relevant topics from the corpus. We then use NLP techniques to generate effective summaries. The produced summaries and topics are evaluated based on their performance, coherence, and interpretability. Our goal is to streamline the process of understanding APIs and methods, enabling developers to extract needed information efficiently.

To guide our research, we aim to answer the following research questions:

\vspace{0.2cm}

\textbf{RQ1}: {What are the recurring topics that developers commonly discuss on Stack Overflow concerning Android APIs?}

\vspace{0.2cm}

\textbf{RQ2}: {Can we employ summarization methods to identify common issues within these topics?}

\vspace{0.2cm}

\textbf{RQ3}: {Is it possible to generate solutions to these common issues by leveraging information from unofficial documentation?}

\section{Related Work}
Recently, the field of automatic summarization has garnered a lot of interest among researchers. The focus, however, has primarily been on summarizing code blocks as opposed to APIs, and only a few studies have utilized unofficial documentation. In this context, we present a review of the most relevant research related to our study, beginning with topic modeling and proceeding to the usage of unofficial documentation.

For instance, Alhaj {\em et al.}~\cite{Alhaj} presented a machine learning-based approach aimed at improving the classification of cognitive distortions in Arabic content shared on Twitter. The proposed method addresses the challenge posed by the shortness of text and the resulting sparsity of co-occurrence patterns and lack of context information. To overcome this, the approach employs BERTopic, which facilitates the enrichment of text representations by uncovering latent topics within tweets. The algorithm utilizes two types of document representations, applying averaging and concatenation techniques to generate contextual topic embeddings. The performance of the approach is evaluated using various metrics such as F1-score, precision, recall, and accuracy. Experimental results indicate that the enriched representation outperforms baseline models by varying degrees. 

Egger {\em et al.}~\cite{Egger} evaluated four topic modeling techniques (LDA, NMF, BERTopic, and Top2Vec) for social science research using Twitter data. BERTopic performs well overall and supports various topic modeling variations. Top2Vec is suitable for large amounts of data but struggles with small datasets. LDA can yield universal or irrelevant information and requires careful hyperparameter tuning. NMF works well with shorter texts like tweets but may result in overlapping clusters. The study suggests that researchers should consider dataset characteristics when choosing a topic modeling technique.

Sridhara {\em et al.}~\cite{Sridhara} have employed Natural Language Processing (NLP) to extract critical information to summarize Java methods. Their findings show that their generated comments accurately represent the source code of Java methods, without losing crucial information during the summarization process.

Abdalkareem {\em et al.}~\cite{Abdalkareem} presented a technique to extract code examples from Stack Overflow posts by utilizing special tags filtering. Their method involves encoding the content of Stack Overflow posts in HTML, enabling the detection of code examples enclosed within the \textless code\textgreater tags through a filtering process. We have adopted this approach in our study since our dataset shared a similar structure, and the code extraction method proved effective in filtering out code examples that were embedded within Stack Overflow posts.


Hu et al.~\cite{Hu} proposed a unique approach to generate code comments from a large codebase using a system called DeepCom. By utilizing NLP techniques, they addressed certain challenges, such as extracting accurate keywords from improperly named methods.

Naghshzan {\em et al.}~\cite{naghshzan} introduced a novel method for creating natural language summaries by leveraging Stack Overflow discussions. The effectiveness of this approach was assessed by surveying 16 developers, who found it to be a valuable additional resource for software development and maintenance tasks. This study makes a contribution to the field of code summarization and underscores the significance of unofficial documentation in the overall process \cite{naghshzan2}.

Bettenburg{\em et al.}~\cite{bettenburg2008extracting} developed InfoZilla, a tool designed to extract various components from bug reports, such as patches, Stack traces, source code, and enumerations. One of the techniques employed by InfoZilla to extract code elements is an island parser inspired by the concept of islands within a sea. This parser identifies islands by searching for specific identifiers. In this context, an "island" refers to classes, conditional statements, functions, and assignments that serve as indicators for locating code elements within discussions. The island parser utilized by InfoZilla builds upon the previous works of Moonen {\em et al.}~\cite{moonen2001generating} and Bacchelli {\em et al.}~\cite{bacchelli2011extracting} in the field of code extraction.

\section{Methodology}

\begin{table*}
  \centering
\caption{Top 10 topics of Android posts in Stack Ovberflow}
\renewcommand{\arraystretch}{1.7}
\begin{tabular}{c|c|l|l}
 \hline
 \textbf{Topic} & \textbf{Count} & \textbf{Name} & \textbf{Representation}\\
 \hline
   1 & 14663 & project\_error\_build\_gradle & project, proguard, studio, error, build, library, file, gradle, android, eclipse \\ \hline
   2 & 13471 & fragment\_viewpager\_view & fragment, recyclerview, item, view, listview, scroll, adapter, list, layout, row \\ \hline
   3 & 9947 & notification\_activity\_service\_gcm & notification, activity, service, gcm, app, analytics, push, back, intent, broadcast \\ \hline
   4 & 7752 & image\_camera\_bitmap\_opencv & image, camera, bitmap, opencv, screen, size, gallery, picasso, wallpaper, picture \\ \hline
   5 & 6474 & bar\_drawer\_navigation\_menu & bar, drawer, navigation, menu, progress, action, seekbar, toolbar, actionbar, layout \\ \hline
   6 & 6425 & keyboard\_search\_edittext\_text & keyboard, search, edittext, text, searchview, input, focus, textview, soft, autocompletetextview \\ \hline
   7 & 5950 & bluetooth\_device\_connection\_connect & bluetooth, device, connection, connect, ble, wifi, usb, android, connected, socket \\ \hline
   8 & 5372 & animation\_tab\_fragment\_viewpager & animation, tab, fragment, viewpager, view, transition, page, animate, activity, pager \\ \hline
   9 & 5320 & play\_google\_purchase\_app & location, gps, sensor, accelerometer, place, get, service, distance, latitude, google \\ \hline
   10 & 4249 & location\_gps\_sensor\_accelerometer & database, sqlite, table, db, query, room, column, cursor, data, insert \\
    \hline
\end{tabular}
  \label{topics}
\end{table*}

Our research methodology consists of three main steps: data collection, topic modeling, and summarization. Each of these steps can be further broken down into individual sub-steps. In the following section, we delve into a comprehensive examination of these key processes.

\subsection{Data Collection and Pre-processing}
We utilized the Stack Exchange API to retrieve all the questions on Stack Overflow that were tagged with {\em Android} from January 2009 to April 2022. Following that, we obtained all the corresponding answers to those questions, resulting in a dataset of 3,698,168. unique Android posts. Since we were interested in natural languages for topic modeling and summarization, we need to remove code blocks from posts. To do this, we used the {\em Code Snippet} feature of Stack Overflow. This feature highlights code segments by wrapping them with an HTML code tag (\textless code\textgreater), making them more accessible to users. By searching for this tag in the posts, we were able to identify code blocks and remove them.

To prepare our data for further analysis, we performed the following pre-processing steps:
\begin{itemize}
    \item Elimination of stop words and punctuation.
    \item Tokenization of text into individual words.
    \item Normalization of words through lemmatization and stemming.
    \item Removal of special characters and numerical values.
\end{itemize}

\subsection{Topic Modeling}
To identify the topics discussed in Stack Overflow, we utilized the BERTopic (Bidirectional Encoder Representations from Transformers Topic Modeling) \cite{bertopic}. BERTopic is a powerful topic modeling technique that leverages the capabilities of pre-trained transformer models, particularly BERT (Bidirectional Encoder Representations from Transformers) \cite{bert}. Traditional topic modeling techniques, such as Latent Dirichlet Allocation (LDA) \cite{Jelodar}, have been widely used to discover latent topics within a corpus of documents. However, these techniques often struggle to capture the semantic meaning and context of words, leading to less accurate and less interpretable topic representations \cite{Egger}. BERTopic overcomes these challenges by integrating the contextual information encoded within BERT's transformer architecture.

To implement BERTopic, we utilized a pre-trained model available on Hugging Face\footnote{https://huggingface.co}. Notably, this model was trained on a vast corpus exceeding one million Wikipedia pages. For the computations, we relied on Google Colab Pro, which comes equipped with a T4 GPU, boasting 25GB of VRAM and 26GB of system RAM. We further leveraged cuML, thereby enabling GPU-accelerated machine learning.

The application of the algorithm resulted in the identification of 1,813 distinct topics. Upon examining the distribution of posts across these topics, we observed that approximately 75\% of the data was concentrated in the top 80 categories. Consequently, we narrowed our research focus to these top 80 topics.

\begin{figure}[!ht]
\centering
\includegraphics[width=\columnwidth]{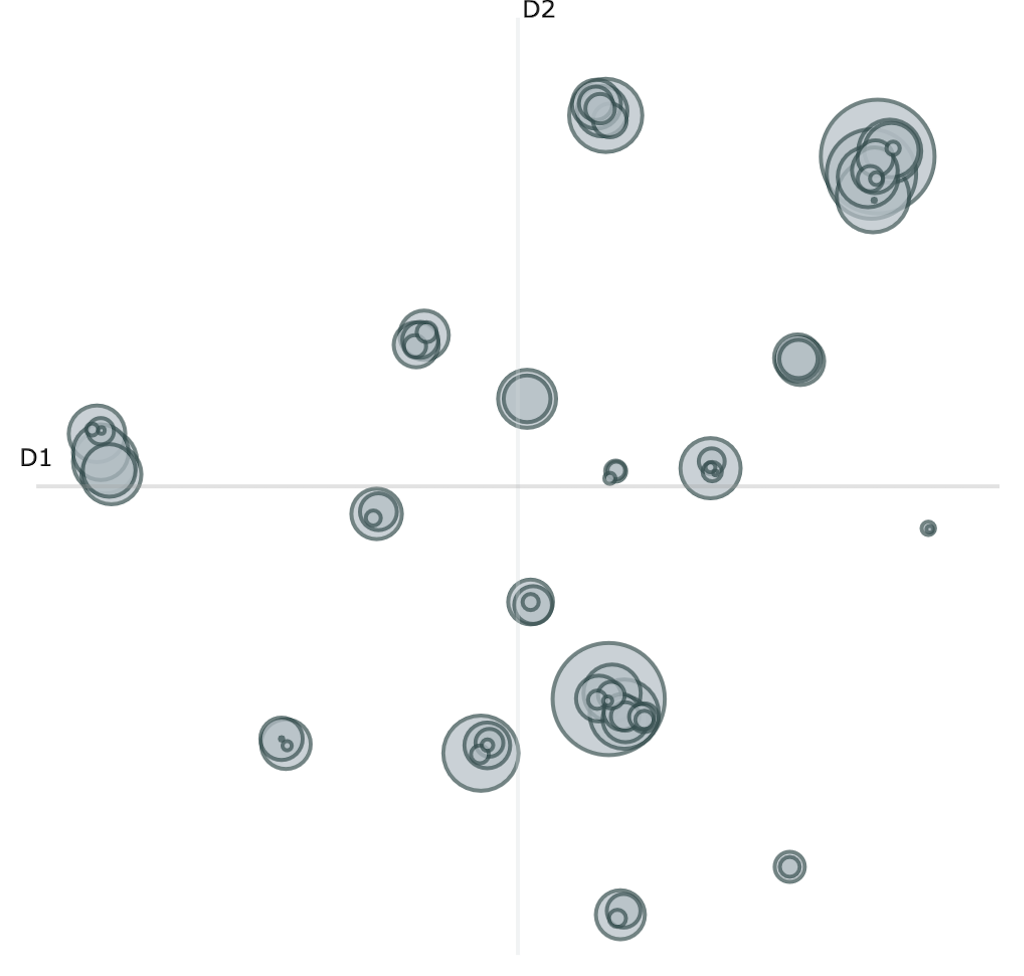}
\caption{Intertopic distance map of top 80 topics}
\label{distance}
\end{figure}

Table \ref{topics} showcases the ten most prevalent topics related to Android posts on Stack Overflow. The \textit{Count} column quantifies the number of posts associated with each specific topic. The \textit{Name} column contains the names of the topics, as generated by BERTopic. The \textit{Representation} column lists the words that best represent each respective topic. This table offers a concise summary of the data, providing insights into the topics that most frequently dominate Android-related discussions on the platform. A comprehensive list of our findings can be accessed in the online Appendix\footnote{https://github.com/scam2023-bert/BERTopic}. Moreover, Figure \ref{distance} visually depicts a two-dimensional distance map of these 80 prominent topics, offering insights into their relational layout and intertopic distances.

\subsection{Summarization}
\begin{table*}
  \centering
\caption{Examples of generated summaries for questions of top two topics.}
\renewcommand{\arraystretch}{1.7}
\begin{tabular}{m{3.5cm} | m{13.55cm}}
 \hline
 \centering{\textbf{Topic Name}}& \multicolumn{1}{c}{\textbf{Summary}} \\
 \hline
 \centering{project\_error\_build\_gradle}   & 
1. jenkins tries to launch tools emulator instead of emulator emulator im trying to set up jenkins ui tests and it fails on running emulator command \vspace{0.12cm} \newline   2. i am trying to add kotlin sources of an aar in android studio it doesnt work when i select choose sources and choose the corresponding source jar \vspace{0.12cm} \newline 3. let me explain the situation i have an sdk works with google api and implementing a lot of dependencies then i implement the lib into my new app which is also implements dependencies everything goes fine until i try to run the app \\
    \hline
 \centering{fragment\_viewpager\_view}  & 
1. hi i am using multiple for loop to add view dynamically in adapter class it causes the application may be doing too much work on its main thread \vspace{0.12cm} \newline  2. error inflating class cardview classnotfoundexception im trying to inflate a layout that contains a cardview but im dealing with some issues \vspace{0.12cm} \newline  3. i thought of setting a listener to each of the layouts for each item but that seems a little too much hassle im sure there is a way for the recyclerview to listen for the onclick event but i cant quite figure it out viewholderjava flatlist pull to refresh not working inside interactable view in android \\
    \hline
\end{tabular}
  \label{questions}
\end{table*}

Our summarization approach comprises two main steps: \\
In the first step, we aim to generate summaries for each topic using questions. The rationale behind this approach is our belief that using questions to summarize topics can aid in identifying the most prevalent problems within each topic. Given that we are focusing on the top topics discussed on Stack Overflow, we hypothesize that we can uncover the most pressing issues in Android development discussed on this platform.

The second step involves generating summaries of topics based on the corresponding answers. These summaries can serve to address the problems identified in the previous step. Therefore, not only do we seek to outline the significant challenges in Android development, but we also strive to provide solutions to these challenges, as derived from the collective wisdom of the Stack Overflow community.

We opted to use BERT as our main algorithm for summarization. BERT (Bidirectional Encoder Representations from Transformers) \cite{bert} is a powerful language model widely used in natural language processing tasks, including text summarization. A common approach to using BERT for text summarization is through the \textit{extractive summarization} approach. In this method, BERT is utilized to assign importance scores to individual sentences or phrases within the input document. The sentences or phrases that receive the highest scores are then chosen and combined to create the summary.

To employ BERT for extractive summarization, the input document is initially tokenized using the BERT tokenizer, which breaks down the text into separate tokens. These tokens are subsequently inputted into the BERT model, which comprises multiple layers of self-attention and feed-forward neural networks. Through this process, the model learns to encode contextual information from the tokens and generate representations that capture the underlying semantic meaning of the text. Once the input document has been encoded, a summarization algorithm based on BERT calculates importance scores for each sentence or phrase. The sentences or phrases that receive the highest scores are selected to form the summary.

As the first step, we employed BERT for generating summaries of questions pertaining to the identified topics. The BERTopic algorithm provided us with the relevant posts representing each topic. Leveraging this functionality, we fed the data into the BERT algorithm to generate summaries. Table \ref{questions} presents a sample of the generated summaries for the top two topics. Each summary consists of three specific problems related to its respective topic. These sentences were assigned the highest BERT scores among all sentences within the topic, indicating that these problems are the most commonly discussed issues on Stack Overflow for that particular topic.

Having identified the top Android problems on Stack Overflow, our next step involves seeking solutions to these issues. To address the identified problems, we perused the posts to locate the related questions in the database. Upon finding a question, we noted its ID. This ID allowed us to find answers associated with that question. Therefore, we employed the same summarization process to create summaries for a question by feeding its answers to our algorithm. To ensure that the generated summary is of sufficient quality, we only selected answers either marked as accepted or with a score above the average for all answers. In this instance, this equates to answers with a score of 2 or higher. 

Table \ref{answers} presents generated summaries of answers for the previously identified problems. Each row in Table \ref{answers} corresponds to an answer for the corresponding problem in Table \ref{questions}. For example, concerning the topic \textit{"project\_error\_build\_gradle"}, one common issue encountered by Android developers is when \textit{"jenkins tries to launch tools emulator instead of emulator"}. Summarizing the answers for this problem suggests a possible solution: \textit{"its an issue with android emulator plugin not working with new command line tools only sdk package"}

\begin{table*}
  \centering
\caption{Examples of generated summaries for answers of top two topics.}
\renewcommand{\arraystretch}{1.7}
\begin{tabular}{m{3.5cm} | m{13.55cm}}
 \hline
 \centering{\textbf{Topic Name}}& \multicolumn{1}{c}{\textbf{Summary}} \\
 \hline
 \centering{project\_error\_build\_gradle}   & 
1. im answering my own question its an issue with android emulator plugin not working with new command line tools only sdk package \vspace{0.15cm} \newline  2. add classpath org jetbrains kotli gradleplugin clean and build your project it worked for me \vspace{0.15cm} \newline 3. the problem is that you have a duplicated dependency with different version if you can find the implementation that needs location you can put under it exclude gms or simply in build gradle module project \\
    \hline
 \centering{fragment\_viewpager\_view}  & 
1. so i solved this problem by changing in manifest file hardwareaccelerated option to true like this \vspace{0.15cm} \newline 2. android studio didnt completely migrate my app to androidx but after changing the package name within the xml layout fixed the issue. \vspace{0.15cm} \newline 3. according to yigit boyar the best way to register a click on a recyclerview is to define the click in the creation of the viewholder instead of just creating a new onclicklistener for each item that the onbindviewholder binds \\
    \hline
\end{tabular}
  \label{answers}
\end{table*}

\section{Discussion}
In this research, we addressed the challenges posed by the extensive and often complex nature of API documentation by introducing a novel approach for generating API summaries from informal documents such as Stack Overflow. Our methodology involved topic modeling using BERTopic, and text summarization using BERT.

To address our first research question: 

\vspace{0.15cm}

\textbf{RQ1}: {What are the recurring topics that developers commonly discuss on Stack Overflow concerning Android APIs?} 

\vspace{0.15cm}

We first identified the recurring topics discussed on Stack Overflow concerning Android APIs. By applying BERTopic modeling to a dataset of 3,698,168 unique Android posts, we extracted 80 prominent topics. These topics provided insights into the most frequently discussed issues in Android development on Stack Overflow.

Moving on to our second research question:

\vspace{0.15cm}

\textbf{RQ2}: {Can we employ summarization methods to identify common issues within these topics?}

\vspace{0.15cm}

We employed text summarization techniques to identify common issues within these topics. By generating summaries for questions related to each topic, we identified the most prevalent problems as derived from the collective wisdom of the Stack Overflow community. These summaries highlighted the key challenges faced by developers in Android programming.

Lastly, addressing our third research question:

\vspace{0.15cm}

\textbf{RQ3}: {Is it possible to generate solutions to these common issues by leveraging information from unofficial documentation?}

\vspace{0.15cm}

We leveraged information from unofficial documentation to generate solutions to the identified common issues. By summarizing answers to the related questions, we provided potential solutions sourced from the knowledge shared by the Stack Overflow community. These solutions aimed to address the challenges identified in the previous step.

In conclusion, our research demonstrated the feasibility of automatically generating API summaries from informal documents like Stack Overflow. By leveraging the power of BERTopic for topic modeling and BERT for text summarization, we were able to extract recurring topics, identify common issues, and generate potential solutions. This approach has the potential to streamline the process of understanding APIs and methods, enabling developers to become familiar with Android development problems and their solutions.

While this study provides valuable insights and lays the foundation for further research, it is important to acknowledge that it represents a preliminary exploration into this area. As such, there is a need for more extensive investigations and studies to validate and expand upon the findings presented here and there is considerable room for improvement and expansion.

\section{Conclusion and Future Work}
In this work, we present a method for automatically generating API summaries from informal sources such as Stack Overflow. By leveraging BERTopic and NLP techniques, we extract recurring topics and create concise summaries of common problems and their solutions. Our approach addresses the challenges faced by developers in extracting relevant information from complex API documentation, providing a streamlined and user-friendly solution. Through the automatic extraction of key insights and their summarization in an accessible manner, our method saves developers valuable time and effort in understanding and utilizing APIs by categorizing common developing problems and their solutions. 

As we continue to refine and expand our approach, there are several avenues for future work that can enhance and expand upon our findings. The following are potential areas for future research: \\
\textbf{Extension to other programming languages and domains:} Although our current study focused on the Android programming language and its APIs, the methodology we presented can be extended to other programming languages and domains. By adapting the techniques to different contexts, we can broaden the applicability and impact of our work. \\
\textbf{Exploration of alternative information sources:} While our research primarily relied on Stack Overflow as the source of informal documentation, there are other platforms and sources of information that developers frequently consult, such as GitHub, blogs, and forums. Exploring these alternative sources and incorporating them into our methodology can provide a more comprehensive and diverse set of data for generating API summaries. \\
\textbf{Investigation of alternative summarization approaches:} While our study utilized BERT-based summarization techniques, there are several other approaches that can be explored to further improve the quality and effectiveness of the generated summaries. Extractive and abstractive summarization methods, as well as hybrid approaches, could be investigated to determine their suitability for API summarization tasks. \\
\textbf{User evaluation studies and feedback gathering:} To assess the utility and effectiveness of the generated API summaries, conducting user evaluation studies and gathering feedback from developers would be invaluable. These evaluations can help identify areas for improvement, validate the usefulness of the summaries in real-world scenarios, and guide further enhancements of the methodology. \\
\textbf{Integration of API summaries into developer tools and IDEs:} An important future direction is the integration of API summaries into developer tools and Integrated Development Environments (IDEs). By seamlessly incorporating the generated summaries within the development workflow, developers can access relevant information more easily, reducing the time spent searching for documentation and improving productivity. 

In conclusion, our method for automatically generating API summaries from informal sources offers potential for aiding developers in understanding and utilizing APIs. By continuing to explore these research directions and addressing the aforementioned areas for future work, we can further enhance the capabilities of our approach and contribute to the advancement of developer productivity and software engineering practices.

\end{document}